\documentstyle[preprint,aps]{revtex} 
\begin{document}
\draft
\title{A Quasi-Spherical Gravitational Wave Solution in Kaluza-Klein Theory}
\author{Malek Zareyan}
\address{Institute for Advanced Studies in Basic Sciences, 
45195-159, Gava Zang, Zanjan, Iran}
\date{\today}
\maketitle 

\begin{abstract}
An exact solution of the source-free Kaluza-Klein field equations is 
presented. It is a 5D generalization of the Robinson-Trautman quasi-spherical
gravitational wave with a cosmological constant. The properties of the 5D 
solution are briefly described.\\
{hep-th1 YY mm nnn}
\end{abstract}
\newpage

\section{Introduction}
The existence of gravitational waves is one of the basic consequences of the 
theory of general relativity. It is also interesting to study gravitational 
radiation within the context of the Kaluza-Klein theory. Exact plane-wave 
solutions of this problem have already been investigated by Liu, Wang and
Shi [1-3]. This work is devoted to the study of quasi-spherical gravitational 
waves in a specific Kaluza-Klein theory. These are solutions to the 5D 
source-free Kaluza-Klein field equations
\begin{equation}
 ^{(5)}R_{AB}=0 \ ,\ \ \ \ (A,B=0,1,2,3,5)\ ,
\end{equation}
with a metric of the form
\begin{equation}
-^{(5)}ds^{2}=\frac{l^2}{L^2}[\frac{r^2}{P^2}d\xi d\bar \xi
-2dudr-(2H-\frac{\Lambda r^2}{3}) 
du^2]+dl^2 . 
\end{equation}
In this metric, $ \xi=x+iy, \bar \xi=x-iy, u$ is 
retarded time, $r$ denotes the affine parameter along the rays of 
gravitational wave, $l$ is the fifth coordinate and $L$ is a constant length
\cite{MLW}. The functions $H=H(\xi,\bar \xi,u,r,l)$ and $P=P(\xi,\bar\xi,u,l)$ 
are to be determined from the field equations (1).\par
The metric (2) is a 5D generalization of the Robinson-Trautman quasi-spherical
gravitational wave with a cosmological constant $\Lambda$. In the 4D 
spacetime, the surfaces $r=u=const$ may be thought of as distorted spheres
(if they  are closed) \cite{MAC}.

\section{ An Exact Solution}
The two unknown functions $H$ and $P$ can be determined from the 5D field 
equations (1). Substituting the metric (2) into Eq. (1), we find 
$$
^{(5)}R_{01}=\frac{1}{P^2} [\Delta  Q-4r Q,_{u}
+\frac{l^2r^2}{L^2}(\frac{P,_{ll}}{P}+\frac{6}{l}Q,_{l})
-\frac{3l^2r^2}{L^2}(Q,_{l})^2-(2rH),_{r}$$
\begin{equation}
+(\Lambda-\frac{3}{L^2})r^2]=0\ ,
\end{equation}
\begin{equation}
^{(5)}R_{02}= Q,_{\xi u }+H,_{\xi r}=0\ ,
\end{equation}
\begin{equation}
^{(5)}R_{05}= Q,_{\xi l}=0\ ,
\end{equation}
\begin{equation}
^{(5)}R_{12}= Q,_{\bar \xi u }+H,_{\bar \xi r}=0\ ,
\end{equation}
\begin{equation}
^{(5)}R_{15}=  Q,_{\bar \xi l}=0\ ,
\end{equation}
$$
^{(5)}R_{22}=(2H-\frac{\Lambda r^2}{3})\ ^{(5)}R_{23}-2P(\frac{1}{P}),_{uu}
+2 Q,_{u}(H,_{r}-\frac{2}{r}H)
+\frac{2}{r^2}P^2H,_{\xi \bar \xi}
$$
\begin{equation} 
+\frac{2}{r}H,_{u}
+\frac{l^2}{L^2}(H,_{ll} 
+\frac{4}{l}H,_{l}-2 Q,_{l}H,_{l})
\end{equation}
\begin{equation}
^{(5)}R_{23}=\frac{2}{r} Q,_{u}- \frac{2l}{L^2}
 Q,_{l}+\frac{1}{r^2}(r^2H,_{r}),_{r}-(\Lambda-\frac{3}{L^2})=0\ ,
\end{equation}
\begin{equation}
^{(5)}R_{25}=-2P(\frac{1}{P}),_{ul}+\frac{1}{r^2}(r^2H,_{l}),_{r} =0\ ,
\end{equation} 
\begin{equation}
^{(5)}R_{35}=\frac{2}{r} Q,_{l}=0\ ,
\end{equation} 
\begin{equation}
^{(5)}R_{55}=\frac{4}{l} Q,_{l}-2P(\frac{1}{P}),_{ll}=0\ ,
\end{equation} 
where $ \Delta =2P^2\partial_{\xi} \partial_ {\bar \xi}$ (which is $r^2/2$ 
times the Laplacian operator on the surface of distorted spheres), 
$Q={\rm ln}P$ and a comma denotes partial differentiation. All other 
components of the 5D Ricci tensor vanish identically. \par   
Note that by virtue of Eq. (11), $P$ is independent of $l$. It then follows 
from $P,_{l}=0$ that $R_{05},R_{15}$ and $R_{55}$ 
vanish identically, and Eq. (3) and Eqs. (8-10) can be written as
\begin{equation}
(2rH),_{r}=\Delta  Q-4r Q,_{u} +(\Lambda-\frac{3}{L^2})r^2\ ,
\end{equation}
\begin{equation}
2P(\frac{1}{P}),_{uu}-2 Q,_{u}(H,_{r}-\frac{2}{r}H)
-\frac{2}{r^2}P^2H,_{\xi \bar \xi}
-\frac{2}{r}H,_{u}
-\frac{l^2}{L^2}(H,_{ll} 
+\frac{4}{l}H,_{l})=0\ ,
\end{equation}
\begin{equation}
2 Q,_{u}+\frac{1}{r}(r^2H,_{r}),_{r}-(\Lambda-\frac{3}{L^2})r=0\ ,
\end{equation}
\begin{equation}
(r^2H,_{l}),_{r}=0\ ,
\end{equation}
respectively. Eqs.(13) and (16) can be integrated, and the results are
\begin{equation}
2H=\Delta  Q-2r Q,_{u}+\frac{1}{3}(\Lambda-\frac{3}{L^2})r^2 
-\frac{2\mu}{r}\ ,
\end{equation}
\begin{equation}
H,_{l}=\frac{f}{r^2}\ ,
\end{equation}
where $f=f(\xi,\bar \xi,u,l)$ and  $\mu=\mu(\xi,\bar \xi ,u,l)$\ are arbitrary 
functions resulting from above integrations. Since $Q$ does not depend on $l$, 
Eqs. (17) and (18) tell us that $f=0$ and so, $H$ and $\mu$ do not depend on 
$l$; moreover, it can be seen from Eq. (4) and Eq. (6) that $\mu$ is a 
function of $u$ alone. Now, using the expression of $H$ from Eq. (17), one can 
verify that Eq. (15) is satisfied identically.\par
Using the above results and substituting Eq. (17) into Eq. (14), we find 
\begin{equation} 
\Delta \Delta Q +12\mu Q,_{u}-4\mu,_{u} =0\ .
\end{equation} 
Eq. (19) corresponds to the familiar equation in the standard 4D 
Robinson-Trautman solution \cite{MAC}. Thus, with $P$ satisfying Eq. (19),we 
obtain an exact gravitational wave solution of the form
\begin{equation}
-\ ^{(5)}ds^{2}=\frac{l^2}{L^2}[\frac{r^2}{P^2}
d\xi d\bar \xi
-2dudr-(\Delta Q-2r Q,_{u}-\frac{2\mu}{r} -\frac{r^2}{L^2}) 
du^2]+dl^2\ ,  
\end{equation}
which satisfies all Eqs. (3-12).\par
Now we impose the condition that the spacetime part of (20),  
\begin{equation}
-\ ^{(4)}ds^{2}=\frac{r^2}{P^2}
d\xi d\bar \xi
-2dudr-(\Delta Q-2r Q,_{u}-\frac{2\mu}{r} -\frac{r^2}{L^2}) 
du^2\ ,  
\end{equation}
should be the gravitational wave solution of the standard 4D Einstein field 
equations with the cosmological term, and from that $\Lambda$ is determine as 
\begin{equation}
\Lambda=\frac{3}{L^2}\ .
\end{equation}
One also may argue that $H$ in (1) should not contain a term proportional to 
$r^2$ and then, using Eq. (17), get the result (22). This is an 
interesting result by itself. It implies that the 5D solution contains the 
parameter $L$ which is related to the cosmological constant in the spacetime.
\par
The solution (20) contains an arbitrary function $\mu(u)$ which by means of
some 4D spacetime coordinate transformation can be chosen to have the value 
$\mu=0$ or $\mu=\pm 1$ \cite{MAC}. A physical interpretation of $\mu$ will be 
worked out by the analysis of the Riemannian curvatures. We have calculated 
all the components of the Riemann tensor for the metric (20), and the 
non-vanishing components are\\
$$
^{(5)}R_{01}\  ^{01}=-2\ ^{(5)}R_{02}\  ^{02}=-2\ ^{(5)}R_{03}\  ^{03}
=-2\ ^{(5)}R_{12}\  ^{12}
=-2\ ^{(5)}R_{13}\  ^{13}=\ ^{(5)}R_{23}\  ^{23}
$$
\begin{equation}
=\frac{2L^2}{l^2}\frac {\mu}{r^3}\ , 
\end{equation}
\begin{equation}
^{(5)}R_{01}\  ^{13}=(\frac{r^2}{P^2})\ ^{(5)}R_{02}\  ^{01}=
-\  ^{(5)}R_{02}\  ^{23}=-(\frac{r^2}{P^2})\ ^{(5)}R_{23}\  ^{13}=
\frac{L^2}{2l^2}\frac{(\Delta  Q),_{\xi}}{r}\ ,
\end{equation}
\begin{equation}
^{(5)}R_{01}\  ^{03}=(\frac{r^2}{P^2})\ ^{(5)}R_{12}\  ^{01}=
\ ^{(5)}R_{12}\  ^{23}=(\frac{r^2}{P^2})\ ^{(5)}R_{23}\  ^{03} 
=-\frac{L^2}{2l^2}\frac{(\Delta  Q),_{\bar \xi}}{r}\ , 
\end{equation}
\begin{equation}
^{(5)}R_{02}\  ^{13}= -\frac{L^2}{l^2}\frac{[P^2(\frac{1}{2}
\Delta  Q-r Q,_{u}),_{\xi}],_{\xi}}{r^2}\ ,
\end{equation}
\begin{equation}
^{(5)}R_{12}\  ^{03}= -\frac{L^2}{l^2}\frac{[P^2(\frac{1}{2}
\Delta  Q-r Q,_{u}),_{\bar \xi}],_{\bar \xi}}{r^2}\ .
\end{equation}
The 5D Kretschmann invariant turns out to be
\begin{equation}
^{(5)}I=\ ^{(5)}R_{ABCD}\ ^{(5)}R^{ABCD}=\frac{48L^4}{l^4}\frac{\mu^2}{r^6}\ .
\end{equation}
The above results indicate that the 5D Riemannian curvature is related to 
$\mu$; all the components of the Riemann tensor are related directly or 
through Eq. (19) to $\mu$.\par 
Solutions of Eq. (20) can be classified on the basis of the values of $\mu$, 
and the associated $P$ functions can be obtained from Eq. (19). For example, 
one of the solutions is characterized by $\mu=0$ and $\Delta  Q=K(u)$ 
\cite{MAC}. In this case, if $P$ is taken to have the special form of
\begin{equation}
P=\alpha (u)\xi \bar \xi+\beta (u)\xi+\bar \beta (u)\bar \xi+\delta (u) \ ,
\end{equation}
with $\alpha, \beta , \bar \beta$  and $\delta$  are arbitrary functions of 
u, all the components of the 5D Riemann tensor will vanish and the 
metric (20) will describe a flat 5D manifold.\par 
Another interesting point is asymptotically flat nature of the 5D manifold in 
the general case; that is, all the components of the Riemann tensor go to 
zero in the limit when the affine parameter $r$ goes to infinity. This can be 
seen easily from Eqs. (23-28).\par
Now let us consider the spacetime hypersurface $ l=const$ and write the 4D 
components of the Riemann tensor. The non-vanishing components are
\begin{equation}
^{(4)}R_{01}\  ^{01}=\ ^{(4)}R_{23}\  ^{23} =2(\frac {\mu}{r^3}
+\frac{\Lambda}{6}) \ ,
\end{equation}
\begin{equation}
\ ^{(4)}R_{02}\  ^{02}=\ ^{(4)}R_{03}\  ^{03}=\ ^{(4)}R_{12}\  ^{12}
 =\ ^{(4)}R_{13}\  ^{13}  =-(\frac {\mu}{r^3}-\frac{\Lambda}
 {3})\ , 
\end{equation}
\begin{equation}
^{(4)}R_{01}\  ^{13}=(\frac{r^2}{P^2})\ ^{(4)}R_{02}\  ^{01}=
-\  ^{(4)}R_{02}\  ^{23}=-(\frac{r^2}{P^2})\ ^{(4)}R_{23}\  ^{13}=
\frac{(\Delta  Q),_{\xi}}{2r}\ ,
\end{equation}
\begin{equation}
^{(4)}R_{01}\  ^{03}=(\frac{r^2}{P^2})\ ^{(4)}R_{12}\  ^{01}=
\ ^{(4)}R_{12}\  ^{23}=(\frac{r^2}{P^2})\ ^{(4)}R_{23}\  ^{03} 
=-\frac{(\Delta  Q),_{\bar \xi}}{2r}\ , 
\end{equation}
\begin{equation}
^{(4)}R_{02}\  ^{13}= -\frac{[P^2(\frac{1}{2}
\Delta  Q-r Q,_{u}),_{\xi}],_{\xi}}{r^2}\ ,
\end{equation}
\begin{equation}
^{(4)}R_{12}\  ^{03}= -\frac{[P^2(\frac{1}{2}
\Delta  Q-r Q,_{u}),_{\bar \xi}],_{\bar \xi}}{r^2}\ ,
\end{equation}
and the 4D Kretschmann invariant is given by
\begin{equation}
^{(4)}I=\ ^{(4)}R_{\mu\nu\rho\sigma}\ ^{(4)}R^{\mu\nu\rho\sigma}
=24(\frac{2\mu^2}{r^6}+\frac{\Lambda^2}{18}),
\end{equation}
The curvature of spacetime depends on parameters $\mu$ and $\Lambda=3/L^2$.
In some special cases, the parameter $\mu$ has the physical meaning of 
the system's mass \cite{MAC}. Here we may interpret $\mu$ as the mass of 
the quasi-spherical gravitational wave. If we take $\mu=0$, the part of the 
spacetime curvature which is related to $\mu$ can be vanish. Then there is 
a curvature in the spacetime that is due to the cosmological constant 
$\Lambda$. This observation shows that the appearance of the cosmological 
constant in the spacetime is somehow related to the contribution of the rest 
of the universe to the field of the quasi-spherical gravitational wave.

\section{Discussion and Conclusion}
The original form  of Kaluza-Klein theory is a 5D theory that attempts to
unify classical gravity and electromagnetism \cite{{KZK},{KLN}}. The question of 
observability of the fifth dimension has been a basic problem of this 
unification \cite{KTH}. An interesting solution to this problem is given in 
the context of a recently developed Machian interpretation of Kaluza-Klein 
theory by Mashhoon, Liu and Wesson \cite{MLW}. This theory is based on a 
classical resolution of the problem of the origin of inertia in Newtonian
mechanics that had been raised by Mach \cite{MCH}.\par 
In the Machian interpretation of Kaluza-Klein theory, the fifth coordinate $l$ 
is directly related to particle mass $m$ through $l=Gm/c^2$; here $G$ is 
Newton's constant and $c$ is the speed of light. This identification of the 
fifth coordinate with mass results in a classical integration of intrinsic and 
extrinsic aspects of a Newtonian point particle in a spacetime-mass manifold.
The spacetime-mass manifold is described by a metric which has a canonical 
form given by
$
\ -\ ^{(5)}ds^{2}=(l^2/L^2)[g_{\alpha\beta}(x,l) dx^{\alpha}dx^{\beta}]
+dl^2, 
$                                                     
in which $L$ is a universal constant. For instance, one may set $L=GM/c^2$, 
where $M$ is considered to be the total mass-energy content of the 
universe. Here $g_{\alpha\beta}$ are the 4D spacetime metric coefficients 
which, in general, depend on the mass coordinate $l$ and the spacetime 
coordinates $x^{\alpha}$ and is determined by the 5D source free field 
equations (1). A particle follows the 5D geodesic equation; hence, the 
particle mass is expected to have a cosmological variation. Moreover, it is 
possible to introduce a constant proper mass for the particle.\par
Eq. (1) is compared with the standard 4D Einstein field equations including 
the cosmological constant term i.e.,
\begin{equation}
 ^{(4)}R_{\mu\nu}-\frac{1}{2}\ ^{(4)}R\ g_{\mu\nu} + \Lambda \ g_{\mu\nu}
 =\frac{8 \pi G}{c^{4}}T_{\mu\nu} \ ,
\end{equation}
where $\Lambda=3/L^2$, and $T_{\mu\nu}$ is an effective 4D energy-momentum 
tensor which represents the induction of matter in spacetime via the fifth 
dimension (it vanishes in the case when $\partial g_{\alpha \beta}/\partial 
l=0$) \cite{{MLW},{LMS}}. \par
The cosmological constant has been interpreted as the average contribution of
the rest of the world to the field of an isolated source, and this 
interpretation has been expected to be true when only a part of the whole is 
taken into account, as would be the case for a gravitational wave \cite{LMS}. 
\par
We have presented an exact quasi-spherical gravitational wave solution of the 
Kaluza-Klein equations, given by the metric (20) and Eq. (19). In accordance 
with the Machian Kaluza-Klein theory, the field equations (2) require that 
the 4D metric, $g_{\alpha\beta}$, be independent of the mass coordinate $l$.
It also follows from these equations that $\Lambda=3/L^2$. A detailed analysis 
of the Riemann curvature tensors reveals the physical significance of the 
cosmological constant. When one considers the construction of 5D 
spacetime-mass manifold, the components of the 5D Riemann tensor depend on 
$\mu$, and the spacetime-mass manifold is expected to be flat for $\mu=0$. 
In the spacetime hypersurface, $l=const$, the rest of the matter in the 
universe contributes an average curvature to the total spacetime curvature via 
the mass coordinate. This contribution appears as the cosmological constant 
and it survives where the part of the curvature, which is related to $\mu$, 
goes to zero ( $r \rightarrow \infty $) Eqs.(30-36). For $\mu=0$, the 
spacetime curvature is expected to be due to the cosmological constant. $\mu$ 
has dimension of length and $(c^2\mu/G)$ can be interpreted as the $proper$ 
$mass$ of the quasi-spherical gravitational wave; that is, $(c^4\mu/G)$ is 
the total energy of the gravitational radiation field.\\

{\large Acknowledgements}\\
I wish to thank Prof. B. Mashhoon for suggesting the problem and fruitful
discussions during this work and also for careful reading the manuscript and 
giving critical comments. I also would like to acknowledge helpful discussions 
with Prof. M. R. H. Khajehpour, Prof. Y. Sobouti and Dr. M. Rainer.
This work was supported by the Institute for Advanced Studies in Basic 
Sciences at Gava Zang, Zanjan, Iran.

\end{document}